\RequirePackage{fix-cm}
\documentclass[smallextended]{svjour3}       
\smartqed  
\usepackage{graphicx}
\begin{document}

\title{Quantum speed limit time of a qubit system with non-Hermitian detuning 
}
\subtitle{}


\author{Yan-Yi Wang$^{1}$ \and
        Mao-Fa Fang$^{1}$$^{\dag}$ 
}


\institute{\at
              $^{1}$Synergetic Innovation Center for Quantum Effects and Applications, and Key Laboratory of Low-Dimensional Quantum Structures and Quantum Control of Ministry of Education, School of Physics and Electronics, Hunan Normal University, Changsha, 410081, China\\
              $^{\dag}$Corresponding author \email{mffang@hunnu.edu.cn}           
}

\date{Received: date / Accepted: date}

\maketitle

\begin{abstract}

We investigated the quantum speed limit time of a qubit system with non-Hermitian detuning. Our results show that, with respect to two distinguishable states of the non-Hermitian system, the evolutionary time does not have a nonzero lower bound. And the quantum evolution of the system can be effectively accelerated by adjusting the non-Hermitian detuning parameter, as well as the quantum speed limit time can be arbitrarily small even be zero.

\end{abstract}

\section{Introduction}
\label{intro}

In conventional quantum mechanics, a Hamiltonian of a quantum system must be represented by a Hermitian operator, which ensures that not only the eigenvalues of the Hamiltonian are real, but also the time evolution of the quantum system is unitary. However, in the last two decades, non-Hermitian systems have received considerable attentions. Many theories of non-Hermitian systems with real and complex spectra have been investigated. In 1998, Bender et~al. proposed that parity-time reversal ($\mathcal{PT}$-) symmetric non-Hermitian Hamiltonians which are invariant under combined space and time reversal still have real and positive energy spectra \cite{non1}, proved that the time-evolution of $\mathcal{CPT}$-symmetry Hamiltonians is unitary, and redefined an inner product whose associated norm is positive definite \cite{non2}, and demonstrated that the evolutionary time of $\mathcal{PT}$-symmetric non-Hermitian systems can even be made arbitrarily small without violating the time-energy uncertainty principle \cite{Faster,optimal}. After Ref. \cite{Faster}, Assis et~al. proved that the phenomenon that the passage time needed for the evolution can be made arbitrarily small in the quantum brachistochrone problem of $\mathcal{PT}$-symmetric systems can also be obtained for non-Hermitian Hamiltonians for which $\mathcal{PT}$-symmetry is completely broken \cite{JPA.259}. G\"{u}nther et~al. demonstrated that the $\mathcal{PT}$-symmetric quantum brachistochrone problem can be reanalyzed as a quantum system consisting of a non-Hermitian $\mathcal{PT}$-symmetric component and a purely Hermitian component simultaneously \cite{paper11261}, and proposed that the quantum mechanical brachistochrone system with a $\mathcal{PT}$-symmetric Hamiltonian can be reinterpreted as a subsystem of a larger conventional quantum mechanics system in a higher-dimensional Hilbert space governed by a Hermitian Hamiltonian \cite{paper1}. Moreover, the framework for the non-Hermitian formalism of Hamiltonians has been proposed by Brody et~al. \cite{NH.mixed} and Sergi et~al. \cite{NH.dynamics}. Basing on the previous works, many quantum properties and quantum effects in non-Hermitian systems have been widely studied \cite{IncPT15,IncPT,NH.time,NH.entanglement3,zhangPT,zhangNH,guo}. These research results show that non-Hermitian Hamiltonians are useful in theoretical work, and they are also regarded as effective mathematical tools for studying quantum properties of open quantum systems in quantum optics \cite{2,open1,open.NHQM,open2.Lindblad}.

On the other hand, the quantum speed limit time originates from the Heisenberg uncertainty relation for energy and time, it is conventionally known as the minimum evolutionary time between two distinguishable states of a system, becomes a key factor in characterizing the maximum evolutionary speed of quantum systems. For closed quantum systems with unitary time evolution, a unified lower bound of the quantum speed limit time is obtained by the Mandelstam-Tamm (MT) type bound $\tau_{QSL}=\pi\hbar/(2\Delta E)$ \cite{MT} and the Margolus-Levitin (ML) type bound $\tau_{QSL}=\pi\hbar/(2E)$ \cite{ML}, where $\Delta E$ is the variance of energy of the initial state and $E$ is the mean energy with respect to the ground state. Both the MT type and the ML type bounds are attainable in closed quantum systems for initial pure states. According to Refs.\cite{optimal,Bender238}, although $\mathcal{PT}$-symmetric non-Hermitian Hamiltonians with real eigenvalues are not Hermitian in the Dirac sense, they do have entirely real spectra and give rise to unitary time evolution. Hence, Ref. \cite{Deffner.JPA6.2} considered that both the MT type bound as well as the ML type bound for time-dependent generators of $\mathcal{PT}$-symmetric non-Hermitian systems to remain valid. However, it should be noted that $\mathcal{PT}$-symmetric Hamiltonians with real eigenvalues and unitary time evolution is a special class of non-Hermitian Hamiltonians. For a general form of non-Hermitian Hamiltonians introduced in Refs. \cite{NH.dynamics} which no longer satisfies $\mathcal{PT}$-symmetric structure called spontaneous $\mathcal{PT}$-symmetry breaking, and the time evolution is non-unitary. Fortunately, Deffner et~al. expressed the quantum speed limit time in terms of the operator norm of the non-unitary generator of the dynamics \cite{Deffner:13}. However, while the unified lower bound in Ref.~\cite{Deffner:13} is applicable for a given driving time for pure initial states, it is not feasible for mixed initial states. Later, Zhang et~al. proposed a generic bound on the evolutionary time of quantum systems with non-unitary time evolution by using the relative purity, which is applicable to both mixed and pure initial states \cite{Zhang:14}. In research field of quantum speed limit time, to effectively reduce the evolutionary time as well as accelerate the speed of quantum evolution is one of our strong expectations. This paper is aimed to investigate the quantum speed limit time in non-Hermitian systems. With the help of numerical calculations, we demonstrate that the non-Hermiticity in Hamiltonians can notably reduce the quantum speed limit time, even decrease to arbitrary small. And we give a potential physical explanation for why the non-Hermitian systems allow for faster evolutions.

This paper is organized as follows. In Sect.~\ref{sec:2}, definitions of the quantum speed limit time are briefly reviewed. In Sect.~\ref{sec:3}, the physical model and the non-Hermitian dynamics are introduced. In Sect.~\ref{sec:4}, the quantum speed limit time of the non-Hermitian quantum system is investigated with the help of numerical calculations. Finally, conclusion and discussion are given in Sect.~\ref{sec:5}.

\section{Definitions of quantum speed limit time}
\label{sec:2}

In this section, we briefly review the definitions of the quantum speed limit time. Deffner and Lutz \cite{Deffner:13} derived a unified lower bound of the quantum speed limit time which is determined by an initial state $\rho_{0}=|\psi_{0}\rangle\langle\psi_{0}|$ and its target state $\rho_{\tau_{D}}$, and governed by arbitrary time-dependent non-unitary equation of the form $\dot{\rho_{t}}=L_{t}\rho_{t}$. With the help of the von Neumann trace inequality and the Cauchy-Schwarz inequality, the quantum speed limit time is obtained as
\begin{equation}
\label{pureQSL}
\tau_{D}\geq\tau_{QSL}=\max\left\{\frac{1}{\Lambda^{1}_{\tau_{D}}},\frac{1}{\Lambda^{2}_{\tau_{D}}},\frac{1}{\Lambda^{\infty}_{\tau_{D}}}\right\}~\sin^2[\mathcal{B}(\rho_{0},\rho_{\tau_{D}})],
\end{equation}
where $\tau_{D}$ is the actual driving time, $\Lambda^{p}_{\tau_{D}}=\tau_{D}^{-1}\int_{0}^{\tau_{D}}\|L_{t}\rho_{t}\|_{p}dt$, and $\|A\|_{p}=(\sigma_{1}^{p}+\cdots+\sigma_{n}^{p})^{1/p}$ denotes the Schatten $p$ norm, $\sigma_{1}, \cdots, \sigma_{n}$ are the singular values of $A$. $\|A\|_{1}=\sum_{i}\sigma_{i}$ is the trace norm, $\|A\|_{2}=\sqrt{\sum_{i}\sigma_{i}^2}$ is the Hilbert-Schmidt norm, and $\|A\|_{\infty}=\sigma_{\max}$ is the operator norm which is given by the largest singular value. Because of the relationship $\|A\|_{\infty}\leq\|A\|_{2}\leq\|A\|_{1}$, the ML-type bound based on the operator norm ($p=\infty$) provides the sharpest bound on the quantum speed limit time. And $\mathcal{B}(\rho_{0},\rho_{\tau_{D}})=\arccos\sqrt{\langle\psi_{0}|\rho_{\tau_{D}}|\psi_{0}\rangle}$ is the Bures angle between the initial state $\rho_{0}$ and its target state $\rho_{\tau_{D}}$. However, Eq.~(\ref{pureQSL}) is not feasible for mixed initial states. Fortunately, Zhang et~al. in Ref.~\cite{Zhang:14} proposed a unified lower bound of the quantum speed limit time for an arbitrary mixed state $\rho_{\tau}$ to its target state $\rho_{\tau+\tau_{D}}$ based on the relative purity:
\begin{equation}
\label{mixQSL}
\tau_{D}\geq\tau_{QSL}=\max\left\{\frac{1}{\overline{\sum^{n}_{i=1}\sigma_{i}\varrho_{i}}},\frac{1}{\overline{\sqrt{\sum^{n}_{i=1}\sigma_{i}^2}}}\right\}*|f(\tau+\tau_{D})-1|~Tr(\rho_{\tau}^2),
\end{equation}
where $\overline{X}=\tau_{D}^{-1}\int_{\tau}^{\tau+\tau_{D}}X dt$, $\sigma_{i}$ and $\varrho_{i}$ are the singular values of $L_{t}\rho_{t}$ and the mixed initial state $\rho_{\tau}$, respectively. And $f(\tau+\tau_{D})=Tr(\rho_{\tau+\tau_{D}}\rho_{\tau})/Tr(\rho_{\tau}^2)$ denotes the relative purity between the initial state $\rho_{\tau}$ and the final state $\rho_{\tau+\tau_{D}}$ with the driving time $\tau_{D}$. For a pure initial state $\rho_{\tau=0}=|\psi_{0}\rangle\langle\psi_{0}|$, the singular value $\varrho_{i}=\delta_{i,1}$, then $\sum^{n}_{i=1}\sigma_{i}\varrho_{i}=\sigma_{1}\leq\sqrt{\sum^{n}_{i=1}\sigma_{i}^{2}}$, and Eq.~(\ref{mixQSL}) can be simplified as
\begin{equation}
\label{mixQSLs}
\tau_{D}\geq\tau_{QSL}=\frac{|f(\tau+\tau_{D})-1|~Tr(\rho_{\tau}^2)}{\sigma_{1}}.
\end{equation}
Eq.~(\ref{mixQSLs}) indicates that, with regard to a pure initial state, the expression given by Eq.~(\ref{mixQSL}) can recover to the unified lower bound of the quantum speed limit time obtained by Eq.~(\ref{pureQSL}). When $\tau_{QSL}=\tau_{D}$, the evolution is already along the fastest path and does not possess potential capacity for further quantum speed-up. While $\tau_{QSL}<\tau_{D}$, the speed-up evolution might occur. And $1/\tau_{QSL}$ defines a natural notion of the speed of quantum evolution. Therefore, reducing $\tau_{QSL}$ would lead to an acceleration of the quantum evolution. Especially, $\tau_{QSL}=0$ can be interpreted as two different situations: for two identical states, $\tau_{QSL}=0$ indicates that the quantum evolutionary speed tends to zero-speed, but for two distinguishable quantum states, $\tau_{QSL}=0$ represents that the quantum evolutionary speed becomes infinity speed.

\section{Physical model and non-Hermitian dynamics}
\label{sec:3}

We all know that the non-Hermitian approach is regarded as one of available methods to describe properties of open quantum systems. Ref.~\cite{NH.dynamics} assumed that in absence of any interaction with the environment, the two-level system is free to make transitions between its two energy levels. Such a situation is modeled by the Hermitian Hamiltonian: $H_{+}$. In order to formulate the open system dynamics of the model, they introduced a general anti-Hermitian Hamiltonian: $H_{-}$. Namely, non-Hermitian Hamiltonians ($H_{nH}\neq H_{nH}^{\dag}$) always can be decomposed into Hermitian and anti-Hermitian parts as $H_{nH}=H_{+}+H_{-}$ with $H_{\pm}=\pm H_{\pm}^{\dag}$, where $H_{-}=-i\Gamma$ can be viewed as the imaginary counterpart of the detuning, and $\Gamma=\Gamma^{\dag}$ can be regarded as the decay rate operator. We choose a special scheme introduced by Ref.~\cite{NH.pure} to describe the non-Hermitian detuning model ($\hbar=1$):
\begin{eqnarray}
\label{H+H-}
H_{+}=-\omega\sigma_{x},~~\Gamma=\gamma\sigma_{z},
\end{eqnarray}
where $\omega$ and $\gamma$ are assumed to be real-valued, $\gamma$ represents the non-Hermitian detuning parameter, and $\sigma_{\alpha}$ ($\alpha=x,z$) are Pauli matrixes. And the Hamiltonian of the non-Hermitian detuning model is given as
\begin{eqnarray}
\label{HnH}
\nonumber H_{nH}&=&-\omega\sigma _{x}-i\gamma\sigma _{z}\\
&=&-\omega\left(
\begin{array}{cccc}
i\Delta&1\\
1&-i\Delta\\
\end{array}
\right),
\end{eqnarray}
where we denoted $\Delta=\gamma/\omega$. In this paper, we set that $\omega^{-1}$ is a scaling constant, namely $\Delta$ has the same sign as the non-Hermitian detuning parameter $\gamma$. And the eigenvalues of $H_{nH}$ are $E_{\pm}=\pm\sqrt{\omega^{2}-\gamma^{2}}=\pm\sqrt{1-\gamma^{2}}$. It is easy to determine that when $\gamma\in(-1,0)\cup(0,1)$, $H_{nH}$ is a $\mathcal{PT}$-symmetric non-Hermitian Hamiltonian with real eigenvalues, while $\gamma\in(-\infty,-1)\cup(1,\infty)$, $H_{nH}$ is a $\mathcal{PT}$-symmetry broken non-Hermitian Hamiltonian with complex eigenvalues. And $\gamma=\pm1$ are usually considered as exceptional points where eigenvalues switch from real values to complex values \cite{Epoints}. When $\gamma=0$, the non-Hermitian detuning degrades into a coherent Rabi oscillation with coupling parameter $\omega=1$. The evolution equation of the non-Hermitian system is non-Schr\"{o}dinger quantum mechanics, but it can be directly derived from Schr\"{o}dinger equation as Refs.~\cite{NH.mixed,NH.dynamics}
\begin{eqnarray}
\label{Omega}
\frac{d}{dt}\Omega_{t}=-i[H_{+},\Omega_{t}]-\{\Gamma,\Omega_{t}\},
\end{eqnarray}
where $[~,]$ and \{~,\} represent the commutator and the anti-commutator, respectively. In general, due to the non-Hermiticity of $H_{nH}$, non-Hermitian dynamics is non-unitary, $\Omega_{t}$ in Eq.~(\ref{Omega}) is a non-normalized density operator. Therefore, for making sure that the density matrix is trace-preserving, the renormalization process is required:
\begin{eqnarray}
\label{rho}
\rho_{t}=\frac{\Omega_{t}}{Tr\Omega_{t}}.
\end{eqnarray}
And then the norm-preserving evolution equation generated by a non-Hermitian Hamiltonian for the normalized density operator is given as
\begin{eqnarray}
\label{dtrho}
\frac{d}{dt}\rho_{t}=-i[H_{+},\rho_{t}]-\{\Gamma,\rho_{t}\}+2Tr(\rho_{t}\Gamma)\rho_{t}.
\end{eqnarray}
It should be noted that the solution of this evolution equation in Eq.~(\ref{dtrho}) also can be expressed in the form as Refs.~\cite{NH.mixed,IncPT}
\begin{eqnarray}
\rho_{t}=\frac{U_{nH}\rho_{0}U_{nH}^{\dag}}{Tr(U_{nH}\rho_{0}U_{nH}^{\dag})},
\end{eqnarray}
where $\rho_{0}$ is the initial state, $U_{nH}=\exp(-i H_{nH}t)$ is still a non-unitary time evolution operator, and the non-Hermitian dynamics is still non-unitary. By solving the evolution equation, we obtain matrix elements $\rho_{t}^{kl}$ ($k,l=1,2$) of the normalized final state $\rho_{t}$ are given as
\begin{eqnarray}
\label{element}
\nonumber\rho_{t}^{11}&=&\frac{1}{\gamma_{1}^{2}T}\{\gamma_{1}^{2}\rho_{0}^{11}\cosh^{2}(\gamma_{1}t)+[1+\gamma_{1}^{2}\rho_{0}^{11}+i \Delta(\rho_{0}^{12}-\rho_{0}^{21})]\sinh^{2}(\gamma_{1}t)\\
\nonumber&&-\gamma_{1}[\Delta\rho_{0}^{11}+\frac{1}{2}i(\rho_{0}^{12}-\rho_{0}^{21})]\sinh(2\gamma_{1}t)\},\\
\nonumber\rho_{t}^{12}&=&\frac{1}{\gamma_{1}^{2}T}[\gamma_{1}^{2}\rho_{0}^{12}\cosh^{2}(\gamma_{1}t)+(i \Delta-\Delta^{2}\rho_{0}^{12}+\rho_{0}^{21})\sinh^{2}(\gamma_{1}t)\\
\nonumber&&+\frac{1}{2}i\gamma_{1}(1-2\rho_{0}^{11})\sinh(2\gamma_{1}t)],\\
\nonumber\rho_{t}^{21}&=&(\rho_{t}^{12})^{*},\\
\rho_{t}^{22}&=&1-\rho_{t}^{11},
\end{eqnarray}
where $\rho_{0}^{ij}$ ($i,j=1,2$) are elements of the initial state $\rho_{0}$, we denoted $T=Tr(U_{nH}\rho_{0}U_{nH}^{\dag})$ and $\gamma_{1}=\sqrt{\Delta^{2}-1}$.

\section{Quantum speed limit time of non-Hermitian detuning model}
\label{sec:4}

According to definitions of the quantum speed limit time given by Eqs.~(\ref{pureQSL})~and~(\ref{mixQSL}), we consider two cases of different initial states $\rho_{0}$.

\subsection{Pure initial state case}

We firstly examine the quantum speed limit time of the qubit system under the non-Hermitian detuning with a pure initial state $|\psi_{0}\rangle=|1\rangle$, and Eq.~(\ref{pureQSL}) can be simplified as
\begin{equation}
\tau_{QSL}=\frac{1-p(\tau_{D})}{\frac{1}{\tau_{D}}\int_{0}^{\tau_{D}}\sigma_{max}dt},
\label{eq11}
\end{equation}
where $p(\tau_{D})=\rho_{\tau_{D}}^{11}$ represents the population of the excited state $|1\rangle$ at the time $\tau_{D}$, and $\sigma_{max}$ is the largest singular value of $L_{t}\rho_{t}$.

\begin{figure}[!htbp]
\centering
  \includegraphics[width=0.6\textwidth]{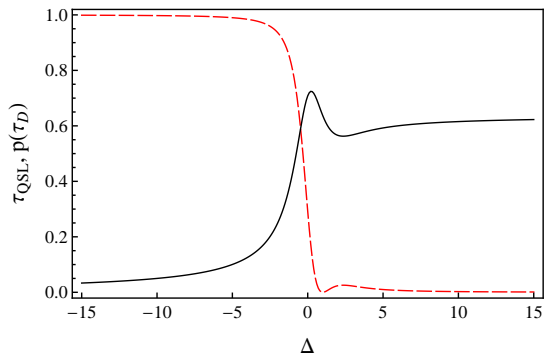}
\caption{The quantum speed limit time $\tau_{QSL}$ (black-solid line) and the excited state population $p(\tau_{D})$ (red-dashed line) as functions of the non-Hermitian detuning parameter $\Delta$ for the initial excited state. The actual driving time is $\tau_{D}=1$.
}
\label{fig:1}
\end{figure}
In Fig.~\ref{fig:1}, we depict the quantum speed limit time $\tau_{QSL}$ (the black-solid curve) as a function of the non-Hermiticity parameter $\Delta$. According to Eq.~(\ref{HnH}) and its explanation, $\Delta$ represents the non-Hermitian detuning parameter, when $\Delta\in(-1,1)$, $H_{nH}$ is a $\mathcal{PT}$-symmetric non-Hermitian Hamiltonian, while $\Delta\in(-\infty,-1)\cup(1,\infty)$, $H_{nH}$ is a $\mathcal{PT}$-symmetry broken non-Hermitian Hamiltonian, and $\Delta=\pm1$ are usually considered as exceptional points. On the basis of definitions of quantum speed limit time, we know that increasing $\tau_{QSL}$ would lead to a deceleration of the quantum evolution, while decreasing $\tau_{QSL}$ would result in an acceleration, and $\tau_{QSL}=0$ can be interpreted as zero-speed or infinity speed. Besides, from Eq.~(\ref{eq11}), one can see that the quantum speed limit time $\tau_{QSL}$ is related to the excited state population $p(\tau_{D})$, and $p(\tau_{D})$ can reflect evolutionary efficiency of the quantum state under the non-Hermitian detuning. In order to explore the internal mechanism of the speed of the quantum evolution, we also plot the excited state population $p(\tau_{D})$ (the red-dashed curve) as a function of the non-Hermiticity parameter $\Delta$. Fig.~\ref{fig:1} indicates that, by adjusting the non-Hermiticity parameter $\Delta$ from negative values to positive values, for negative and small region $\Delta\in(-15,-3)$, the excited state population $p(\tau_{D})$ slightly decreases, and the quantum speed limit time $\tau_{QSL}$ increases correspondingly. Especially, when $\Delta\in(-3,1.5)$, $p(\tau_{D})$ quickly decreases, and $\tau_{QSL}$ firstly increases, and then decreases. For positive and large region $\Delta\in(1.5,15)$, $p(\tau_{D})$ slightly reverts, and finally decreases to zero, $\tau_{QSL}$ firstly decreases and then increases to a constant. That is to say, when $\Delta$ in negative and small region, although the value of $\tau_{QSL}$ is smaller, $p(\tau_{D})$ barely changes, which means that the evolution of the quantum state is faster but low efficient. In the region which $p(\tau_{D})$ rapidly and efficiently changes, the quantum evolution experiences a process from deceleration to acceleration, and then $\tau_{QSL}$ gradually tends to a constant speed with $\Delta$ increases continuously. It is obvious that, with regard to the non-Hermiticity parameter $\Delta$, the excited state population $p(\tau_{D})$ has a negative relation with the quantum speed limit time $\tau_{QSL}$.\\

\begin{figure}[!htbp]
  \includegraphics[width=0.5\textwidth]{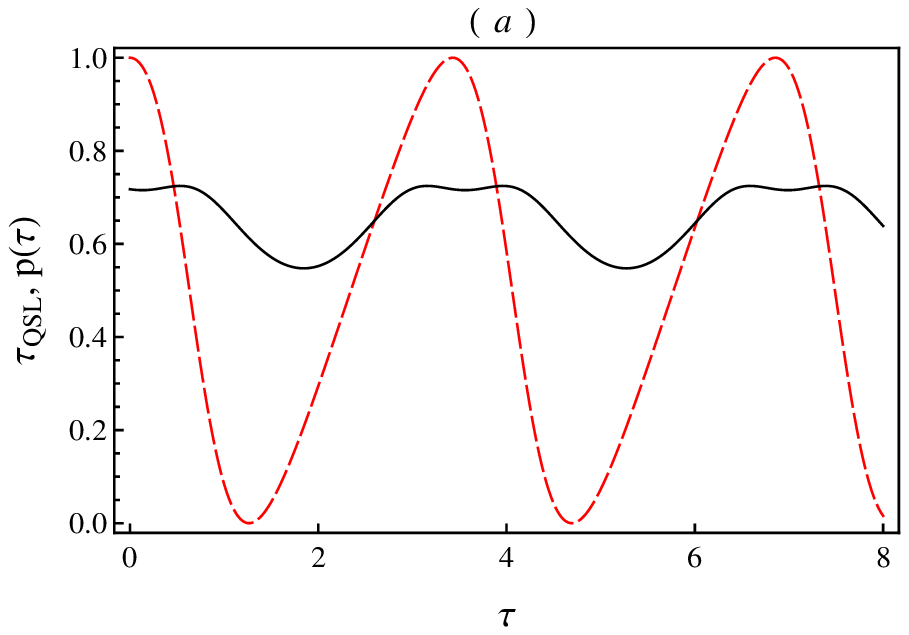}
  \includegraphics[width=0.5\textwidth]{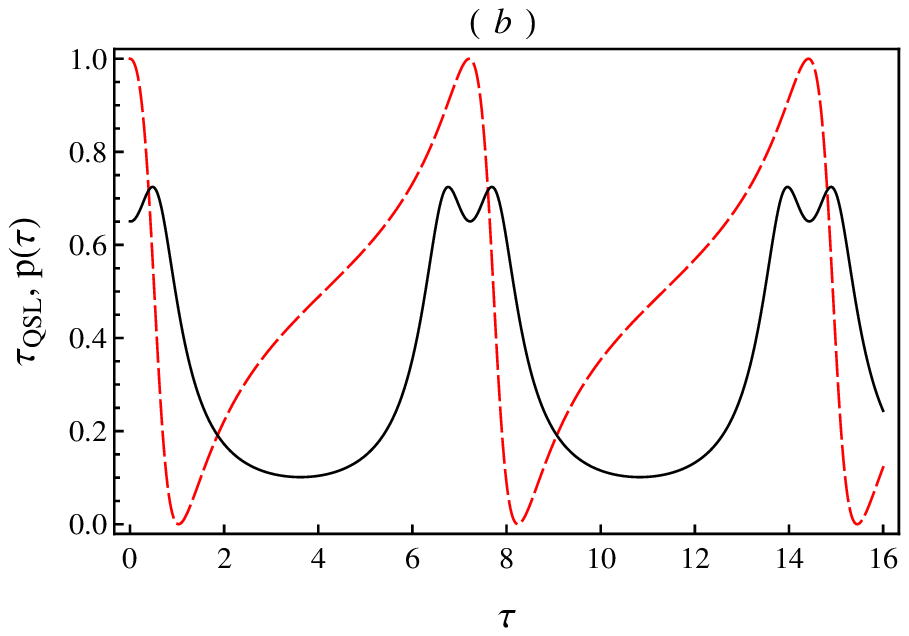}
\caption{The quantum speed limit time $\tau_{QSL}$ (black-solid line) and the excited state population $p(\tau)$ (red-dashed line) as functions of the initial time parameter $\tau$ for the initial excited state. (a) $\Delta=0.4$; (b) $\Delta=0.9$. The actual driving time $\tau_{D}=1$.
}
\label{fig:2}
\end{figure}

\begin{figure}[!htbp]
  \includegraphics[width=0.5\textwidth]{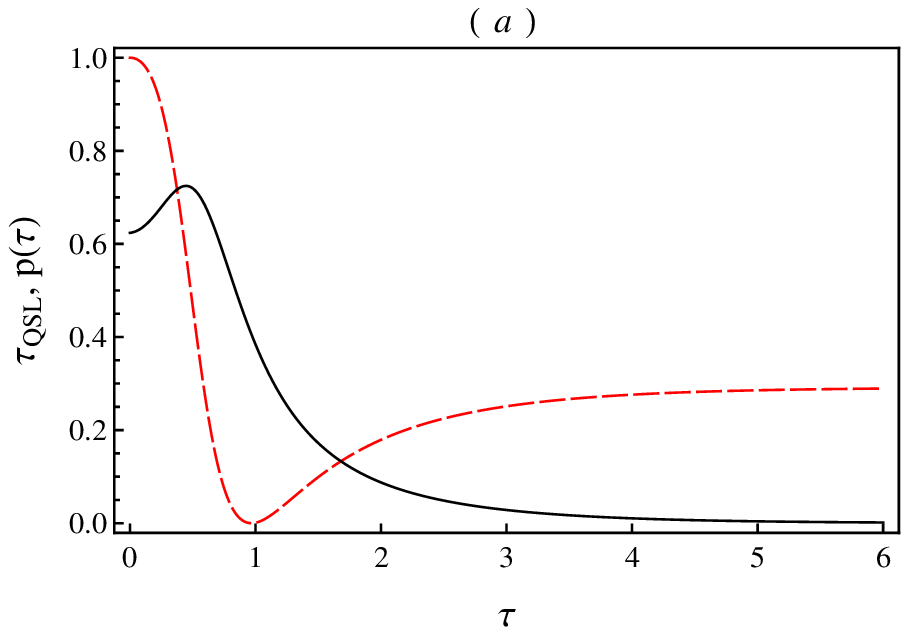}
  \includegraphics[width=0.5\textwidth]{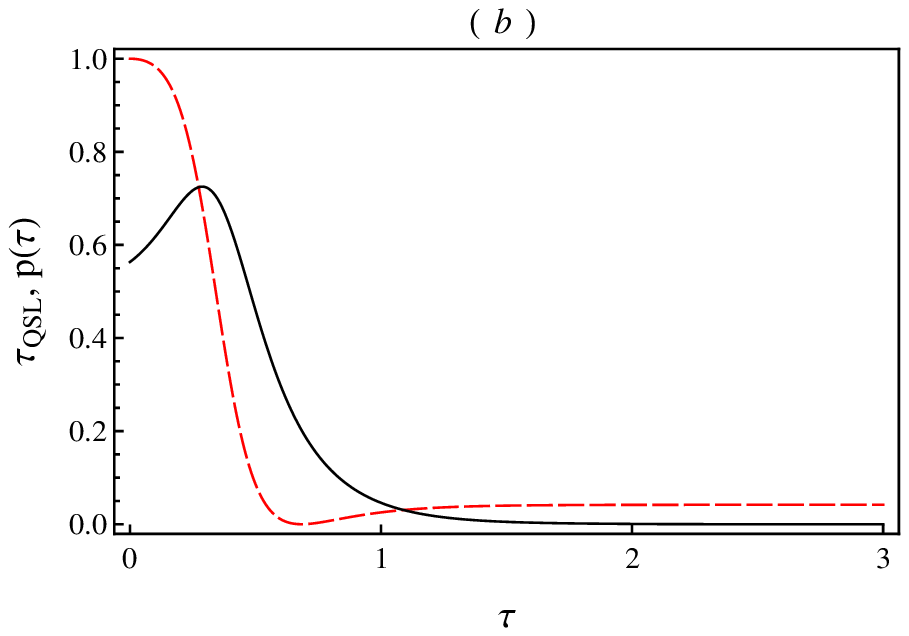}
\caption{The quantum speed limit time $\tau_{QSL}$ (black-solid line) and the excited state population $p(\tau)$ (red-dashed line) as functions of the initial time parameter $\tau$ for the initial excited state. (a) $\Delta=1.1$; (b) $\Delta=2.5$. The actual driving time $\tau_{D}=1$.
}
\label{fig:3}
\end{figure}

We secondly explore effects of the non-Hermitian detuning on the quantum speed limit time in the whole dynamical process using Eq.~(\ref{mixQSL}). We also start from the excited state $|1\rangle$ and discuss the quantum speed limit time from an arbitrary state $\rho_{\tau}$ to its target state $\rho_{\tau+\tau_{D}}$, and $p(\tau)=\rho_{\tau}^{11}$ represents the population of the excited state $|1\rangle$ at the time $\tau$. In Figs.~\ref{fig:2}~and~\ref{fig:3}, we plot the quantum speed limit time $\tau_{QSL}$ (black-solid curves) and the excited state population $p(\tau)$ (red-dashed curves) as functions of the initial time parameter $\tau$ for different non-Hermiticity parameter $\Delta$. According to Eq.~(\ref{HnH}) and its explanation, we know that the non-Hermitian Hamiltonian with the non-Hermiticity parameter $\Delta=$~0.4~or~0.9 in Fig.\ref{fig:2} satisfies $\mathcal{PT}$-symmetric structure. In Fig.\ref{fig:2}, the quantum speed limit time $\tau_{QSL}$ periodically decreases and increases, which implies that the quantum evolution of the whole dynamical process exists periodical speed-up and speed-down. We also find that with $\Delta$ increases from 0.4 to 0.9, the effect of acceleration and deceleration becomes more obvious. It is means that, with $\mathcal{PT}$-symmetric, the larger the non-Hermiticity parameter $\Delta$ is, the more obvious the effect of speed-down and speed-up is, the smaller value the quantum speed limit time can be achieved, which corresponds to a faster speed of the quantum evolution. And the excited state population $p(\tau)$ periodically evolves with the initial time parameter $\tau$, and has the same period with the quantum speed limit time $\tau_{QSL}$. While as the non-Hermiticity parameter $\Delta=$~1.1~or~2.5 in Fig.~\ref{fig:3}, the non-Hermitian Hamiltonian does not satisfy $\mathcal{PT}$-symmetric structure, and the quantum evolution is aperiodic. From Fig.~\ref{fig:3}, we can see that the quantum speed limit time $\tau_{QSL}$ slightly increases at the beginning, and gradually decreases to zero, which implies that the quantum evolution in the whole dynamical process exists a speed-down and a speed-up. And the excited state population $p(\tau)$ firstly decreases and then increases, finally reaches at a stable value. The zero quantum speed limit time combined with a stable excited state population indicates that the state of the system finally evolves into a steady state and the quantum evolutionary speed finally tends to zero-speed. And the larger the non-Hermiticity parameter $\Delta$ is, the faster the evolution to the steady state is, the smaller the stable value of the excited state population $p(\tau)$ is, which corresponds to a faster and more effective quantum evolution.

\subsection{Mixed initial state case}

In this subsection, we examine the quantum speed limit time of the qubit system under the non-Hermitian detuning with a mixed initial state
\begin{eqnarray}
\label{mixed}
\rho_{0}=(1-\frac{p}{2})|1\rangle\langle1|+\frac{p}{2}|0\rangle\langle0|,~~(0<p<1)
\end{eqnarray}
where $p$ is a constant parameter. For clearly demonstrating the quantum evolution process, the trace distance is also considered. The trace distance is a measure of the distinguishability between two quantum states $\rho_{1}$ and $\rho_{2}$ \cite{distance}:
\begin{eqnarray}
D(\rho_{1},\rho_{2})=\frac{1}{2}Tr|\rho_{1}-\rho_{2}|,
\end{eqnarray}
with $Tr|A|=Tr\sqrt{A^{\dag}A}$. The trace distance of two distinguishable states satisfies the inequality $0<D(\rho_{1},\rho_{2})<1$, $D(\rho_{1},\rho_{2})=0$ for two identical states $\rho_{1}=\rho_{2}$, and $D(\rho_{1},\rho_{2})=1$ for two orthogonal states $\rho_{1}\rho_{2}=0$.

\begin{figure}[!htbp]
  \includegraphics[width=0.5\textwidth]{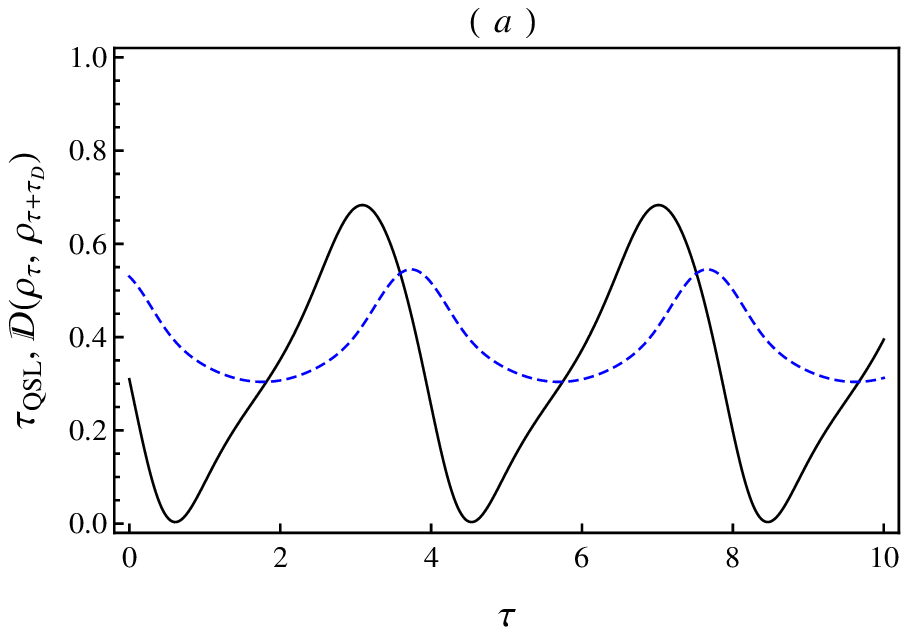}
  \includegraphics[width=0.5\textwidth]{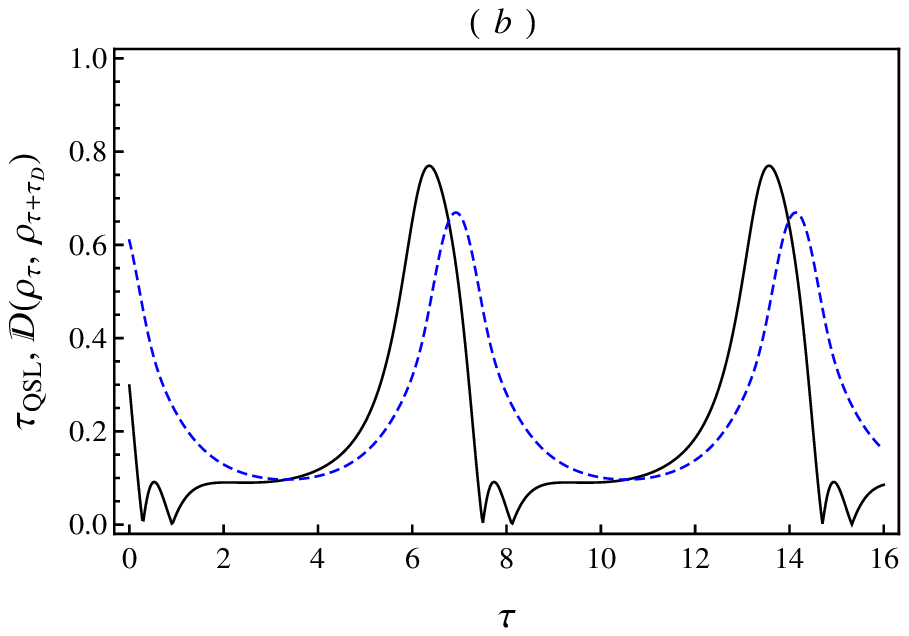}
\caption{The quantum speed limit time $\tau_{QSL}$ (black-solid line) and the trace distance $D(\rho_{\tau},\rho_{\tau+\tau_{D}})$ (blue-dotted line) as functions of the initial time parameter $\tau$  for a mixed initial state. (a) $\Delta=0.6$; (b) $\Delta=0.9$. Other parameters are $\tau_{D}=1$ and $p=0.6$.
}
\label{fig:4}
\end{figure}
\begin{figure}[!htbp]
  \includegraphics[width=0.5\textwidth]{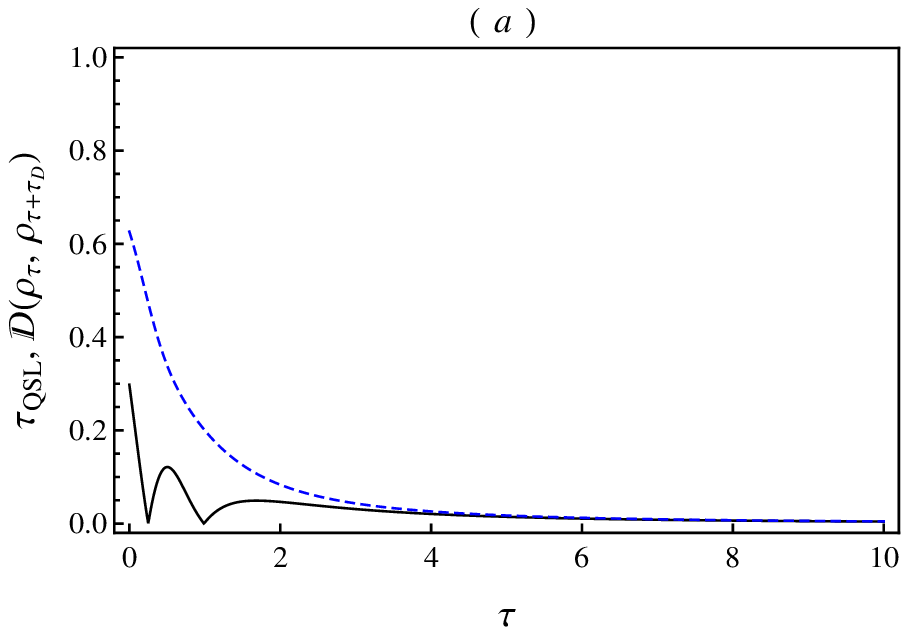}
  \includegraphics[width=0.5\textwidth]{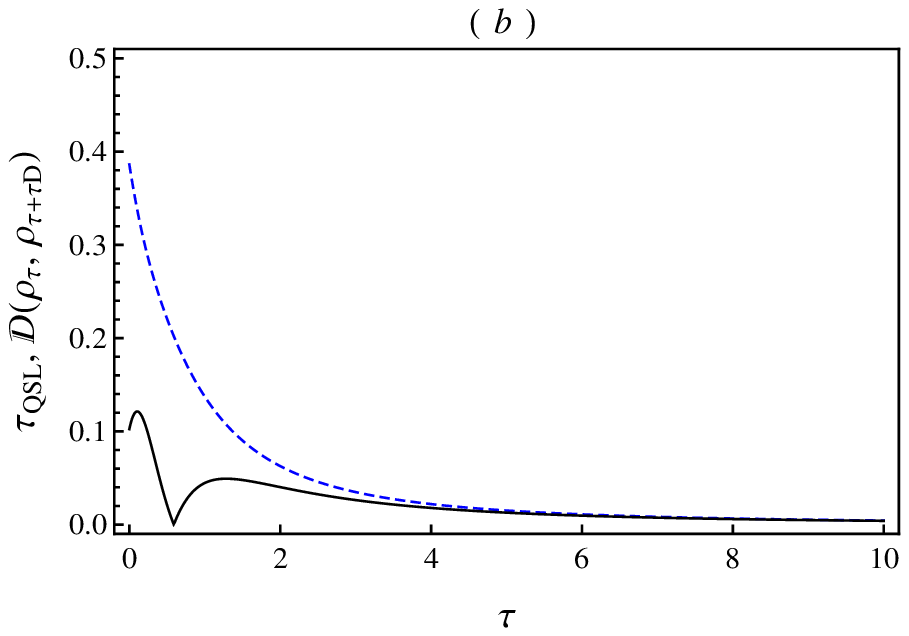}
\caption{The quantum speed limit time $\tau_{QSL}$ (black-solid line) and the trace distance $D(\rho_{\tau},\rho_{\tau+\tau_{D}})$ (blue-dotted line) as functions of the initial time parameter $\tau$  for a mixed initial state. (a) $\Delta=1$; (b) $\Delta=-1$. Other parameters are $\tau_{D}=1$ and $p=0.6$.
}
\label{fig:5}
\end{figure}
Effects of the non-Hermitian detuning on the quantum speed limit time in the whole dynamical process can be studied using Eq.~(\ref{mixQSL}) when the initial state takes the form of general mixed state given by Eq.~(\ref{mixed}). In Figs.~\ref{fig:4}~and~\ref{fig:5}, we depict the quantum speed limit time $\tau_{QSL}$ (black-solid curves) and the trace distance $D(\rho_{\tau},\rho_{\tau+\tau_{D}})$ (blue-dotted curves) as functions of the initial time parameter $\tau$ for different non-Hermiticity parameter $\Delta$. According to Eq.~(\ref{HnH}) and its explanation, we know that the non-Hermitian Hamiltonian with the non-Hermiticity parameter $\Delta=$~0.6~or~0.9 in Fig.~\ref{fig:4} satisfies $\mathcal{PT}$-symmetric structure, and the quantum evolution is periodical. We find that with the non-Hermitian detuning parameter $\Delta$ increases from 0.6~to 0.9, the acceleration and deceleration of the quantum evolution of the whole dynamical process becomes more multiple. In Fig.~\ref{fig:4}~(a), $\Delta=$~0.6, the quantum evolution exhibits straightforward acceleration and deceleration in one period. While in Fig.~\ref{fig:4}~(b), $\Delta=$~0.9, the quantum speed limit time $\tau_{QSL}$ tends to zero twice in one period, and the quantum evolution experiences multiple speed-down and speed-up. It is worth pointing out that when the quantum speed limit time $\tau_{QSL}$ reduces to a minimum, the trace distance $D(\rho_{\tau},\rho_{\tau+\tau_{D}})$ is nonzero. In Fig.~\ref{fig:5}, non-Hermiticity parameters $\Delta=$~$1$~and~$-1$ are exceptional points of $\mathcal{PT}$-symmetric structure, and the quantum evolution is aperiodic. We can also find that when the quantum speed limit time $\tau_{QSL}$ decreases to zero, the trace distance $D(\rho_{\tau},\rho_{\tau+\tau_{D}})$ is still nonzero as same as Fig.~\ref{fig:4}. The nonzero trace distance and the zero quantum speed limit time indicates that the quantum evolutionary speed tends to infinity speed. That is to say, the evolutionary time of non-Hermitian systems do not have a nonzero lower bound, which is a remarkable difference comparing with the traditional quantum theory. In addition, we also consider situations that $\Delta$ is negative and $\Delta>1$. Our numerical calculations show the same result that the quantum evolutionary speed of distinguishable states tends to infinity speed under the non-Hermitian detuning. However, an exceptional point of the non-Hermitian Hamiltonian has only one (geometric) eigenvector, since both the eigenvalues and the corresponding eigenstates of the non-Hermitian system coalesce at exceptional points, Ref.~\cite{JPA250} found that some non-Hermitian Hamiltonian corresponds to a non-Hermitian degeneracy called exceptional points, and showed that any state evolution can be generated solely by such non-Hermitian degeneracies yielding an exceptional points-driven evolution which minimizes the Hilbert-Schmidt norm of the matrix of non-Hermitian Hamiltonians. Hence, we choose some typical and representative cases to expound our work.\\

As mentioned in Sect.~\ref{sec:4}, numerical results show that the quantum speed limit time of the quantum evolution between two distinguishable states of non-Hermitian systems can tend to zero. And this result also satisfies the surprising result in Refs.~\cite{Faster,JPA.259} which demonstrated that the evolutionary time of non-Hermitian systems can be made arbitrary small without violating the time-energy uncertainty principle.

\section{Conclusion and discussion }
\label{sec:5}

In this paper, we considered a qubit system with non-Hermitian detuning, and studied the quantum speed limit time of the non-Hermitian system in regard to two cases of pure and mixed initial states. In the pure initial state case, the quantum evolution of the system can be effectively accelerated by adjusting the non-Hermitian detuning parameter. While in the mixed initial state case, with respect to two distinguishable states of the non-Hermitian system, that the evolutionary time does not have a nonzero lower bound, as well as the quantum speed limit time can be arbitrarily small even be zero. In summary, non-Hermitian systems can be regard as good candidates for achieving ultrafast quantum evolution.

According to other's research results of previous works, we give a probable physical explanation for why the non-Hermitian systems allow for faster evolutions. As we all know that, in conventional quantum mechanics, the Hamiltonian of a physical system requires the Hermiticity to ensure that the energy of the system is real and that the time evolution of the system is unitary. However, a quantum system described by a non-Hermitian Hamiltonian possesses the non-Hermiticity, which usually leads to that the time evolution of the system is a non-unitary evolution. In order to give the non-Hermitian system a meaning in conventional quantum mechanics, Ref.~\cite{non1} proved that the major different between conventional and non-Hermitian quantum mechanics is the definition of the inner product, and constructed an inner product for the non-Hermitian quantum mechanics called the $\mathcal{CPT}$ inner product whose associated norm is positive definite. Because the Hilbert-space metric depends on the Hamiltonian, the geometry of Hilbert space of the non-Hermitian quantum theory has to be modified. Hence, a pair of states is orthogonal under the standard inner product of the Hermitian quantum theory, but is no longer orthogonal under the $\mathcal{CPT}$ inner product of the non-Hermitian quantum theory. As a consequence, in Ref.~\cite{Faster}, Bender noted that it is possible to create a wormholelike effect in the Hilbert space to explain why the transformation between a pair of orthogonal states (under the standard inner product in Hermitian quantum theory) can be made in arbitrarily small time. This is because for non-Hermitian Hamiltonians the alternative complex pathway from a state to its orthogonal state can be made arbitrarily short. The mechanism described here is similar to that in general relativity in which the alternative distance between two widely separated space-time points can be made small if they are connected by a wormhole. Besides, Ref.~\cite{paper1} proposed that the non-Hermitian system can be reinterpreted as a subsystem of a Hermitian system in a higher-dimensional Hilbert space, and the embedding of non-Hermitian system into a higher-dimensional Hilbert space can be deemed as a strengthening of the wormhole analogy introduced by Ref.~\cite{Faster}. As mention above, because of the $\mathcal{CPT}$ inner product, the geometry of Hilbert space of the non-Hermitian quantum theory has to be modified. It is easy to note that, the modification of non-Hemitian Hilbert space has to be implemented, irrespective of whether two distinguishable states of non-Hermitian systems are orthogonal or not. That is to say, under the $\mathcal{CPT}$ inner product of the non-Hermitian quantum theory, by adjusting the non-Hermiticity of the system, the alternative complex pathway of such two distinguishable states can be made in arbitrarily short, and the evolutionary time of the states also can be arbitrarily small without violating the time-energy uncertainty principle.

\begin{acknowledgements}
This work is supported by the National Natural Science Foundation of China (Grant No.11374096).
\end{acknowledgements}


\begin{thebibliography}{}
%
%

\bibitem{non1}
Carl M. Bender, and Stefan Boettcher, Real Spectra in Non-Hermitian Hamiltonians Having PT Symmetry, Phys. Rev. Lett., 80(24), 5243 (1998).

\bibitem{non2}
Carl M. Bender, Dorje C. Brody, and Hugh F. Jones, Complex Extension of Quantum Mechanics, Phys. Rev. Lett., 89, 270401 (2002).

\bibitem{Faster}
Carl M. Bender, Dorje C. Brody, Hugh F. Jones, and Bernhard K. Meister, Faster than Hermitian Quantum Mechanics, Phys. Rev. Lett., 98, 040403 (2007).

\bibitem{optimal}
Carl M. Bender, and Dorje C. Brody, Optimal Time Evolution for Hermitian and Non-Hermitian Hamiltonians, Lect. Notes Phys., 789, 341-361 (2009).

\bibitem{JPA.259}
Paulo E G Assis, and Andreas Fring, The quantum brachistochrone problem for non-Hermitian Hamiltonians, J. Phys. A: Math. Theor., 41, 244002 (2008).

\bibitem{paper11261}
Uwe G\"{u}nther, and Boris F. Samsonov, PT-symmetric brachistochrone problem, Lorentz boosts, and nonunitary operator equivalence classes, Phys. Rev. A, 78, 042115 (2008).

\bibitem{paper1}
Uwe G\"{u}nther, and Boris F. Samsonov, Naimark-Dilated PT-Symmetric Brachistochrone, Phys. Rev. Lett., 101, 230404 (2008).

\bibitem{NH.mixed}
Dorje C. Brody, and Eva-Maria Graefe, Mixed-State Evolution in the Presence of Gain and Loss, Phys. Rev. Lett., 109, 230405 (2012).

\bibitem{NH.dynamics}
Alessandro Sergi, and Konstantin G. Zloshchastiev, Non-Hermitian Quantum Dynamics of a Two-Level System and Models of Dissipative Environments, Int. J. Mod. Phys. B, 27(27), 1350163 (2013).

\bibitem{IncPT15}
Yi-Chan Lee, Min-Hsiu Hsieh, Steven T. Flammia, and Ray-Kuang Lee, Local PT symmetry violates the no-signaling principle, Phys. Rev. Lett., 112, 130404 (2014).

\bibitem{IncPT}
Shin-Liang Chen, Guang-Yin Chen, and Yueh-Nan Chen, Increase of entanglement by local PT-symmetric operations, Phys. Rev. A, 90, 054301 (2014).

\bibitem{NH.time}
Alessandro Sergi, and Konstantin G. Zloshchastiev, Time correlation functions for non-Hermitian quantum systems, Phys. Rev. A, 91, 062108 (2015).

\bibitem{NH.entanglement3}
C. Li, and Z. Song, Generation of Bell, W, and Greenberger-Horne-Zeilinger states via exceptional points in non-Hermitian quantum spin systems, Phys. Rev. A, 91, 062104 (2015).

\bibitem{zhangPT}
Zhang Shi-Yang, Fang Mao-Fa, Zhang Yan-Liang, Guo You-Neng, Zhao Yan-Jun, and Tang Wu-Wei, Reduction of entropic uncertainty in entangled qubits system by local PT-symmetric operation, Chin. Phys. B, 24(9), 090304 (2015).

\bibitem{zhangNH}
Shi-Yang Zhang, Mao-Fa Fang, and Lan Xu, Quantum entropy of non-Hermitian entangled systems, Quantum Inf. Process, 16, 234 (2017).

\bibitem{guo}
You-neng Guo, Mao-fa Fang, Guo-you Wang, Jiang Hang, Ke Zeng, Enhancing parameter estimation precision by non-Hermitian operator process, Quantum Inf. Process, 16, 301 (2017).

\bibitem{2}
M. B. Plenio ,and P. L. Knight, The quantum-jump approach to dissipative dynamics in quantum optics, Rev. Mod. Phys., 70(1), 101 (1998).

\bibitem{open1}
Ingrid Rotter, A non-Hermitian Hamilton operator and the physics of open quantum systems, J. Phys. A: Math. Theor., 42, 153001 (2009).

  \bibitem{open.NHQM}
Nimrod Moiseyev, Non-Hermitian Quantum Mechanics, Cambridge University Press, Cambridge (2011).

\bibitem{open2.Lindblad}
Konstantin G. Zloshchastiev, and Alessandro Sergi, Comparison and unification of non-Hermitian and Lindblad approaches with applications to open quantum optical systems, J. Mod. Opt., 61(16), 1298-1308 (2014).

\bibitem{MT}
L. Mandelstam, and Ig. Tamm, The uncertainty relation between energy and time in nonrelativistic quantum mechanics, J. Phys. (USSR), 9, 249-254 (1945).

\bibitem{ML}
Norman Margolus, and Lev B. Levitin, The maximum speed of dynamical evolution, Phys. D, 120, 188-195 (1998).

\bibitem{Bender238}
Carl M Bender, Making sense of non-Hermitian Hamiltonians, Rep. Prog. Phys., 70, 947-1018 (2007).

\bibitem{Deffner.JPA6.2}
Sebastian Deffner, and Steve Campbell, Quantum speed limits: from Heisenberg's uncertainty principle to optimal quantum control, J. Phys. A: Math. Theor., 50, 453001 (2017).

\bibitem{Deffner:13}
Sebastian Deffner, and Eric Lutz, Quantum Speed Limit for Non-Markovian Dynamics, Phys. Rev. Lett., 111, 010402 (2013).

\bibitem{Zhang:14}
Ying-Jie Zhang, Wei Han, Yun-Jie Xia, Jun-Peng Cao, and Heng Fan, Quantum speed limit for arbitrary initial states, Sci. Rep., 4, 4890 (2014).

\bibitem{NH.pure}
Konstantin G. Zloshchastiev, Non-Hermitian Hamiltonians and stability of pure states, Eur. Phys. J. D, 69, 253 (2015).

\bibitem{Epoints}
W D Heiss, Exceptional points of non-Hermitian operators, J. Phys. A: Math. Gen., 37, 2455 (2004).

  \bibitem{distance}
M. A. Nielsen and I. L. Chuang, Quantum Computation and Quantum Information: 10th Anniversary Edition. Cambridge University Press, New York (2011).

\bibitem{JPA250}
Raam Uzdin, Uwe G\"{u}nther, Saar Rahav, and Nimrod Moiseyev, Time-dependent Hamiltonians with 100\% evolution speed efficiency, J. Phys. A: Math. Theor., 45, 415304 (2012).

\end{thebibliography}


\end{document}